\documentclass[11pt]{article}
\usepackage{graphicx}
\newcommand{\BABARPubYear}    {04}

\newcommand{\BABARProcNumber} {101}
\newcommand{\SLACPubNumber} {10807}

\input pubboard/babarsym
\def\bztopippim{\ensuremath{\Bz \to \pip \pim}\xspace}
\def\bptopippiz{\ensuremath{\Bp \to \pip \piz}\xspace}
\def\bztopizpiz{\ensuremath{\Bz \to \piz \piz}\xspace}
\def\bptorhoppiz{\ensuremath{\Bpm \to \rho^{\pm} \piz}\xspace}

\def\bptoKppiz{\ensuremath{\Bp \to \Kp \piz}\xspace}
\def\bztoKzpiz{\ensuremath{\Bz \to \Kz \piz}\xspace}
\def\bztoKzpip{\ensuremath{\Bp \to \Kz \pip}\xspace}

\def\bptoKpKz{\ensuremath{\Bp \to \Kp \Kzb}\xspace}
\def\bztoKzKz{\ensuremath{\Bz \to \Kz \Kzb}\xspace}
\newcommand{\css}[3]{\ensuremath{#1 \pm #2 \pm #3}}
\newcommand{\asym}[1]{\ensuremath{{\cal A}_{#1}}}
\newcommand{\asymd}[2]{\ensuremath{{\cal A}_{#1}^{#2}}}
\def\cpipi{\ensuremath{{C_{\pi\pi}}}}
\def\spipi{\ensuremath{{S_{\pi\pi}}}}
\def\deltae{\ensuremath{\Delta E}}
\def\fish{\ensuremath{{\cal F}}}
\def\etal{et al.}
\def\BR{\ensuremath{{\cal B}}}

\begin{document}
{\pagestyle{empty}

\begin{flushright}
SLAC-PUB-\SLACPubNumber \\
\babar-PROC-\BABARPubYear/\BABARProcNumber \\
\end{flushright}

\par\vskip 4cm

\begin{center}
\Large \bf
Charmless 2- and 3-body B decays and the angle $\alpha$($\phi_2$)
\end{center}
\bigskip

\begin{center}
\large 
Markus Cristinziani\\
(representing the \lbabar\ collaboration)\\
Stanford Linear Accelerator Center\\
Menlo Park, CA 94025\\
E-mail: markus@slac.stanford.edu
\end{center}
\bigskip \bigskip

\begin{center}
\large \bf Abstract
\end{center}
We present preliminary
measurements of branching fractions and \CP-asymmetry
parameters in two- and three-body charmless hadronic \B decays. The
available data sample consists of $227$ million
$\Y4S\to\BB$ decays collected with the \babar\ detector at the \pep2\ asymmetric-energy
$\epem$ collider at SLAC. We establish the observation of the decays $\bztopizpiz$ and $\bztoKzKz$
and constrain the CKM angle $\alpha$ with a full SU(2) isospin analysis in the
$\B \to \pi\pi$ system and with a $\Bz \to \pip \pim \piz$ time-dependent Dalitz plot analysis.
\vfill
\begin{center}
Contributed to the Proceedings of the 32$^{nd}$ International 
Conference on High Energy Physics,\\ 
ICHEP'04, 16 August --- 22 August 2004, Beijing, China
\end{center}

\vspace{1.0cm}
\begin{center}
{\em Stanford Linear Accelerator Center, Stanford University, 
Stanford, CA 94309} \\ \vspace{0.1cm}\hrule\vspace{0.1cm}
Work supported in part by Department of Energy contract DE-AC02-76SF00515.
\end{center}
\section{Introduction}

According to the Standard Model \CP\ violation is attributed to
the presence of one complex phase in the CKM quark-mixing matrix.
The relations between the matrix elements $V_{ij}$ are usually
represented as a triangle in the complex plane, the Unitarity Triangle. The program of the
\B factories aims at overconstraining 
its sides and angles.
Most measurements of branching fractions and \CP\ parameters presented
in this talk can be used to extract information about the
angle $\alpha = \arg\left[-V_{\rm td}^{}V_{\rm tb}^{*}/V_{\rm
    ud}^{}V_{\rm ub}^{*}\right]$.

More detail about the analyses presented here can be found in the
conference contributions\cite{all}.

\section{Hadronic Charmless \B Decays}

These results are based on the analysis of
$227$ million \BB decays recorded by the
\babar\ detector at the \pep2\ asymmetric-energy
\epem\ collider at SLAC. 
The \babar\ detector is described in detail elsewhere\cite{nim}.

Decays of a \B meson into final states with two or three charmless
particles are rare, with branching fractions typically of 
${\cal{O}}(10^{-5})$.
Signal decays are identified using two kinematic variables: (1) the
difference \deltae\ between the energy of the \B candidate
in the \epem center-of-mass (CM) frame and $\sqrt{s}/2$ and (2) the beam-energy
substituted mass
$\mes = \sqrt{(s/2 + {\mathbf {p}}_i\cdot {\mathbf {p}}_B)^2/E_i^2- {\mathbf {p}}_B^2}$,
where $\sqrt{s}$ is the total CM energy, and the B momentum ${\mathbf {p_B}}$
and the four-momentum of the initial state $(E_i, {\mathbf {p_i}})$ are
defined in the laboratory frame.

The main common background consists of continuum ($\epem \to \qqbar$)
events where two or three mesons combine kinematically to mimic
a \B decay. To suppress this jet-like background, a cut
on the sphericity of the event is applied. 
Additionally, a Fisher discriminant \fish\ is defined 
as an optimized linear combination of $\sum_i
p_i$ and $\sum_i p_i \cos^2{\theta_i}$,
where $p_{i}$ is the momentum and $\theta_{i}$ is the angle with
respect to the thrust axis of the \B candidate, both in the CM frame,
for all tracks and neutral clusters not used to reconstruct the \B
meson.
Alternatively a neural network is trained on those two variables and the
angles with respect to the beam axis of the \B momentum and \B thrust
axis in the $\Y4S$ frame.
Background sources from \B decays are vector-pseudoscalar decays, where one of the decay 
products remains undetected, and cross-feed among the charmless modes.

The determination of \CP\ parameters relies on the tagging technique and
a precise measurement of the flight time.
Those particles in the event that are not used to reconstruct
the decay mode under study provide information about whether the other \B
meson decayed as a \Bz or \Bzb.
The \CP\ asymmetry parameters in $\Bz\to\pip\pim$ decays are determined with a
maximum likelihood fit including information about the \B flavor and the
difference $\deltat$ between the decay times.
The decay rate distribution $f_+\,(f_-)$ for the tagged $\B = \Bz\,(\Bzb)$ is given by

\[
f_{\pm}(\Delta t) = \frac{e^{-\left|\deltat\right|/\tau}}{4\tau} 
[1 \pm \spipi\sin(\deltamd\deltat) \mp \cpipi\cos(\deltamd\deltat)],
\]

where $\tau$ is the mean $\Bz$ lifetime and $\deltamd$ is the mixing
frequency due to the neutral-$B$-meson eigenstate mass difference.

All new results described here are summarized in the two tables showing
branching fractions and \CP\ parameters.

\subsection{$B \to \pi \pi$ modes}

We updated the time-dependent \CP\ asymmetry measurement in the 
decay \bztopippim. After selection of events with two charged tracks,
a maximum-likelihood fit is performed using \mes, \deltae, \fish\ and $\theta_C$,
the \v{C}erenkov angle measured by the detector of internally reflected \v{C}erenkov light
which provides good $K-\pi$ separation in the relevant momentum region. 
Signal and background yields of the four related $h^+ h^-$ 
modes ($h \equiv \pi,K$) are determined in a first fit and fixed in 
the final fit where information about \B-flavor and decay-time is added.
We measure the \CP\ parameters in the decay \bztopippim to be
$\cpipi = -0.09 \pm 0.15 \pm 0.04$ and $\spipi = -0.30 \pm 0.17 \pm 0.03$
which does not indicate presence of significant \CP\ violation. 
As shown in Fig.~\ref{fig:pipi} this result is not compatible with Belle's
measurement with 152 million \Bz's\cite{belle}.

For the analysis of the modes \bptopippiz and \bztopizpiz candidate
\piz mesons are reconstructed as pair of photons in the electromagnetic
calorimeter with requirements on minimum energy and lateral shower shape.
For high momentum \piz's the two-photon mass resolution is approximately 8 \mevcc.
For both the \bztopizpiz signal and the \bptorhoppiz
background the \mes and \deltae\ variables are correlated and therefore a
two-dimensional PDF from a smoothed, simulated distribution is used.
To eliminate systematic uncertainties associated with the choice of 
fit function of the \fish\ distribution, a parametric
step function is used\cite{pi0pi0}.
The result of the maximum likelihood fit for \bztopizpiz is
$n(\bztopizpiz) = 61 \pm 17$.  
The significance of the event yield
is found to exceed $5.0\sigma$ including systematic effects. 
The event yield is transformed into a measurement of the branching fraction
$\BR(\bztopizpiz) = ( 1.17 \pm 0.32 \pm
0.10 )\times 10^{-6}$.
Considering the improved understanding of the \piz detection efficiency and
the additional data this result is consistent with our previous measurement\cite{pi0pi0}.
In the same fit the
time-integrated \CP\ asymmetry, defined as 
$C_{\piz\piz} = (\left|A_{00}\right|^2 - \left|\overline{A}_{00}\right|^2)/(\left|A_{00}\right|^2 + \left|\overline{A}_{00}\right|^2)$,
where $A_{00}$ ($\overline{A}_{00}$) is the
{\ensuremath{\Bz (\Bzb) \to \piz\piz}\xspace}
decay amplitude is measured. We find $C_{\piz\piz}= -0.12 \pm 0.56 \pm 0.06 $.
Finally the charge asymmetry and branching fraction for the
decay \bptopippiz are measured and shown in the tables.

\subsection{\mbox{Twobody charmless decays with kaons}}

$\B \to K \pi$ decays are dominated by $b \to s$ penguin transitions 
and are interesting modes to look for possible new physics or constrain
the CKM angle $\gamma$\cite{zoltan}.
New results presented here are included in the tables.
We note that the charge asymmetry $\asym{\Kp \piz} = (6 \pm 6 \pm 1)\%$
is consistent with zero,
while the measured direct asymmetry 
$\asym{\Kp \pim} = (-13.3 \pm 3.0 \pm 0.9)\%$ is not\cite{jinwei}.
The time-dependent \CP\ parameters of 
$\B \to \KS \piz$ are related to the angle $\beta$ and discussed in\cite{hoecker}.

The branching fraction and asymmetry of the previously unobserved 
decay $\Bz \to \Kz \Kzb$ is measured with a significance of $4.5\sigma$
including systematic uncertainties. 
Figure~\ref{fig:K0K0} shows the background-subtracted \deltae\ distributions.
The background subtraction is performed by weighting events using the
$_{s}\cal P$$lot$ technique\cite{pivk}.

\subsection{$\Bz \to \rho^\pm \pi^\mp$}
The final state of the decay $\Bz \to \rho^\pm \pi^\mp$ is not a 
\CP\ eigenstate and the decay $\Bz \to \rho^0 \pi^0$ has not yet
been observed. A direct extraction of $\alpha$ using 
simple isospin relations like in the $\B \to \pi \pi$ system 
does not appear promising.
Instead, we performed a full time-dependent 
Dalitz analysis of the charmless
three-body system $\Bz \to \pip \pim \piz$
with 213 million \BB pairs,
which allows a theoretically cleaner extraction of the angle $\alpha$\cite{sny} 
compared to the previously adopted quasi-twobody approach.

The 16 coefficients of the bilinear form factor terms occurring
in the time-dependent decay rate of the \Bz meson are determined 
in a maximum-likelihood fit with an event yield of 
$n(\Bz \to \pip \pim \piz) = 1184 \pm 58$.
The physically relevant quantities
are derived from these coefficients, resulting in the measurement
of the direct \CP-violation
$\asym{\rho \pi}=-0.088\pm 0.049\pm{0.013}$ and
$C=0.34\pm 0.11\pm{0.05}$ and the mixing-induced \CP-violation parameter
$S=-0.10\pm 0.14 \pm{0.04}$. For the dilution and strong phase shift 
we obtain $\Delta C=0.15\pm 0.11 \pm{0.03}$
and $\Delta S=0.22\pm 0.15\pm{0.03}$, respectively.
These results can be expressed in terms of the asymmetries 
\asymd{\rho\pi}{+-} (\asymd{\rho\pi}{-+}), which involve only
diagrams where the $\rho (\pi)$ meson is emitted by the W boson,
and are shown in Tab.~2.
For the relative strong phase $\delta_{+-}$
between the $\Bz\to\rho^-\pi^+$ and $\Bz\to\rho^+\pi^-$ transitions we find
$(-67^{\,+28}_{\,-31}\pm7)^\circ$.

\section{Extraction of $\alpha$}

We use the isospin relations of reference\cite{Isospin} to
extract information on the angle difference $\delta = \alpha-\alpha_{\rm eff}$,
based on the measurement of the branching fraction\cite{babarpipi} 
$\BR(\bztopippim) = (4.7\pm 0.6\pm 0.2)\times 10^{-6}$
in conjunction with the asymmetries $C_{\pi^+\pi^-}$ and $C_{\piz\piz}$
and the \bztopizpiz and \Btopipiz decay rates described here.
We scan over all values of $\left| \delta \right|$
and calculate a $\chi^2$ for the decay amplitudes,
given these five measurements and the two isospin constraints for each value of
$\left| \delta \right|$. The $\chi^2$ is converted into a confidence level, as shown in Fig.~\ref{fig:pi0pi0},
from which we derive an upper bound on  $\left|\delta\right|$  of $35^{\rm o}$ at the 90\% C.L.

From the measured coefficients of the amplitude relations in the 
Dalitz analysis we can extract an independent bound on $\alpha$, with little
theoretical assumptions. We find $\alpha = \left(113^{\,+27}_{\,-17}\pm6\right)^\circ$,
while only a weak constraint is achieved at the significance level of
more than two standard deviations.

\section*{Acknowledgments}
I would like to thank the organizers for an enjoyable and stimulating conference, and
my \babar\ colleagues for their assistance and helpful discussions.


\clearpage

\begin{table}
\begin{center}
\begin{tabular}{|c|c|c|}
\hline
Decay & ${\cal B} \times 10^{-6}$ & $N\sigma$\\
\hline\hline
\bptopippiz & \css{5.8}{0.6}{0.4} & \\
\bztopizpiz & \css{1.17}{0.32}{0.10} & $5.0$\\
\hline
\bptoKppiz &  \css{12.0}{0.7}{0.6} & \\
\bztoKzpiz &  \css{11.4}{0.9}{0.6} & \\
\bztoKzpip &  \css{26.0}{1.3}{1.0} & \\
\hline
\bztoKzKz & \css{1.19}{0.38}{0.13} & $4.5$ \\
\bptoKpKz & $< 2.35\ \ \  90\%$ C.L. &\\
\hline
\end{tabular}
\caption{Summary of branching fractions measured with 227 million \BB pairs. The last
column ($N\sigma$) shows the significance including systematic effects.}\label{tab:smtab}
\end{center}
\end{table}

\begin{table}
\begin{center}
\begin{tabular}{|c|c|}
\hline
\ \ Parameter\ \ & \hspace{14mm} Value \hspace{14mm} \\
\hline\hline
\spipi & \css{-0.30}{0.17}{0.03}\\
\cpipi & \css{-0.09}{0.15}{0.04}\\
\asym{\pip \piz} & \css{-0.01}{0.10}{0.02} \\
$C_{\ppz}$ & \css{-0.12}{0.56}{0.06} \\
\hline
\asym{\Kp \piz} & \css{0.06}{0.06}{0.01} \\
\asym{\Kz \pip} & \css{-0.087}{0.046}{0.010} \\
\hline
\asymd{\rho\pi}{+-} & \css{-0.21}{0.11}{0.04}\\
\asymd{\rho\pi}{-+} & \css{-0.47}{0.15}{0.06}\\
\hline
\end{tabular}
\caption{Summary of updated \CP\ parameters.}\label{tab:smtab2}
\end{center}
\end{table}

\begin{figure}[!tbp]
\begin{center}
\includegraphics[width=0.6\linewidth]{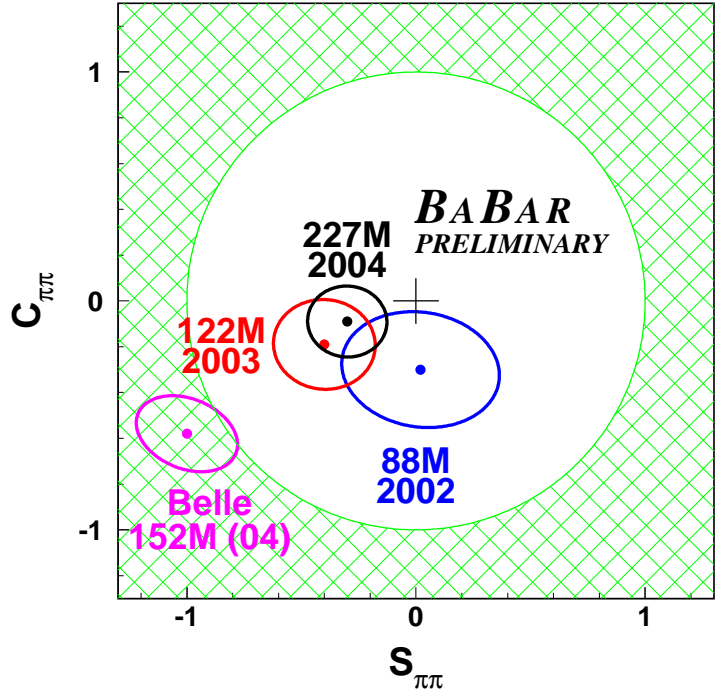}
\end{center}
\caption{Central values and $1\sigma$ contours of the time-dependent \CP\ parameters
\cpipi\ and \spipi\ in the
decay \bztopippim on different \babar\
datasets in contrast to the measurement from Belle.}
\label{fig:pipi}
\end{figure}

\begin{figure}[!tbp]
\begin{center}
\includegraphics[width=0.5\linewidth]{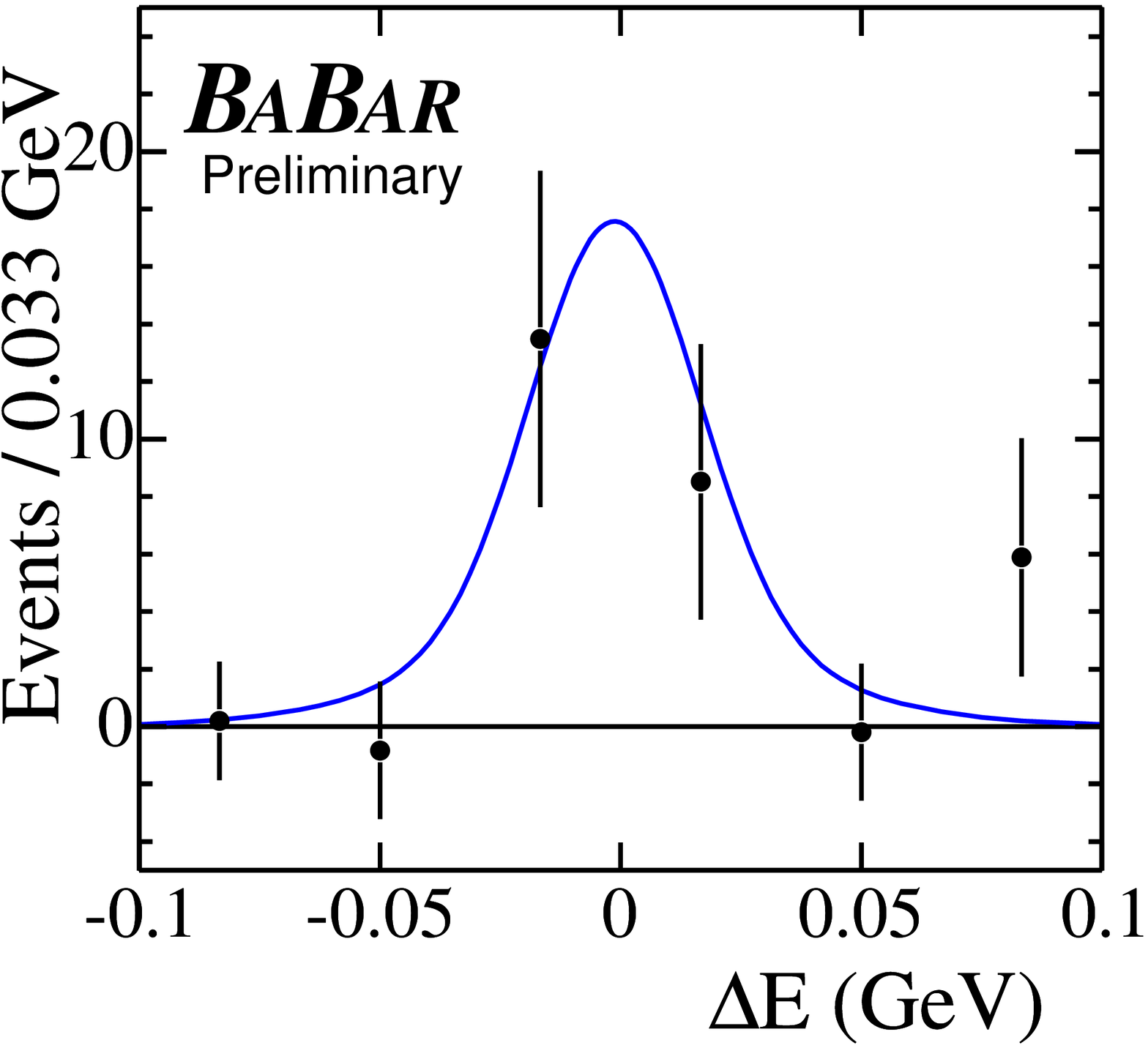}
\end{center}
\caption{\deltae\ distribution for background subtracted \bztoKzKz events (see text). }
\label{fig:K0K0}
\end{figure}

\begin{figure}[!tbp]
\begin{center}
\includegraphics[width=0.6\linewidth]{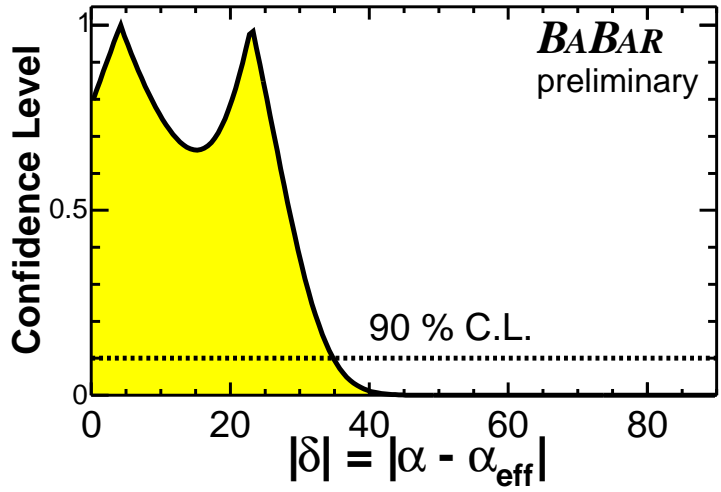}
\end{center}
\caption{Confidence level for the parameter $\delta$ from the full
$B \to \pi \pi$ isospin analysis.}
\label{fig:pi0pi0}
\end{figure}

\begin{figure}[!tbp]
\begin{center}
\includegraphics[width=0.6\linewidth]{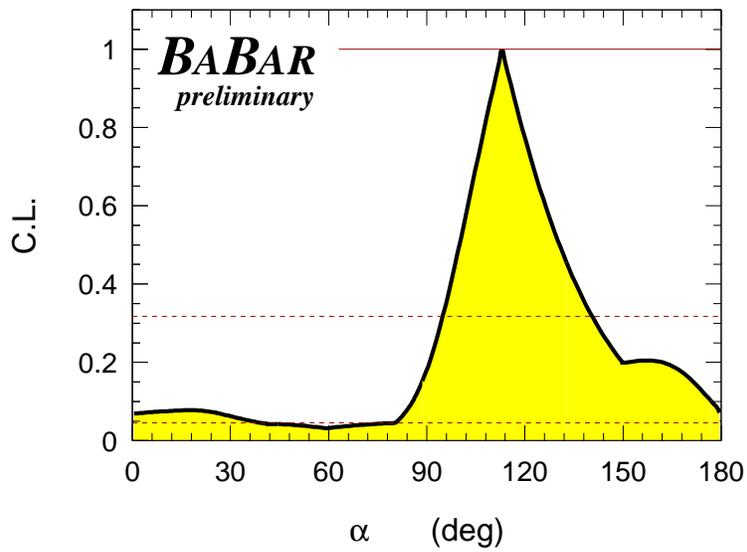}
\end{center}
\caption{Confidence level for the CKM angle $\alpha$ from the $\Bz \to \pip \pim \piz$ Dalitz analysis.}
\label{fig:rhopi0}
\end{figure}

\end{document}